\documentstyle[din_a4,12pt]{article}
\begin{document}
\begin{center}
\Large
{\bf A Statistical Treatment to the Evolution Equation for Quark Distributions  } \\
\large \vspace{1.0cm}
Hai Lin \\
\small {June, 2001}
\\
 \vspace{0.5cm}
\normalsize
 $School$ $of$ $ Physics,$ $ Peking $ $University,$ $ P.R.China,$ $ 100871$\\
{$ Email:hailin@mail.phy.pku.edu.cn$} \\
\vspace{0.5cm}
\end{center}
\begin{abstract}
The evolution of polarized quark distribution functions is taken into account the
gluon emission and absorption, quark pair production and annihilation processes and
treated by a statistical method which provides quark distribution functions at small
$Q^2$. The evolution function of quark distributions is derived and solved by
numerical methods. Then the $Q^2$ dependent and $\alpha_{s}$ dependent structure
functions $F_{2}(x, Q^{2})$ are obtained and the example of it, the proton structure
function, is calculated, whose increase and decrease tendency with $\alpha_{s}$ at
different $x$ is consistent with experiments. This is also an evidence that
the statistical methods for the nucleon is reasonable.

\vspace{-0.3cm}
\end{abstract}

\vspace{1.0cm}
\normalsize

Following our statistical model proposed in [1, 2], we would like to extend our
discussion to the $Q^{2}$ dependence of the quark distribution functions, but only
one part of the evolution equations in QCD. In the early discussion, we didn't take
into account the detailed processes involving gluons, for example, gluon emission and
absorption by quarks. By a combination of effective field theory [1, 3, 4, 5] and
statistical methods [2, 6, 7, 8, 9, 10, 11, 12], the quark distribution at small
$Q^{2}$ is obtained as follows [2],
\begin{equation}
q(x)=\frac{3M^{2}Rx}{2\pi}{\int}_{\frac{Mx}{2}}^{\frac{M}{2}} \frac{1}
{\exp(\frac{E-\mu}{T})+1} \frac{dE}{E^{2}},
\end{equation}
where, $T, R, M$ are the nucleon's temperature, radius and mass, $\mu$ is the quark's
potential.

When $Q^{2}$ is big, in the e-p deep inelastic scattering, the electron can probe
many other detailed processes other than $\gamma^{*}q\rightarrow q$ [13, 14, 15, 16,
17]. We assume that the dominant detailed process involves a photon and the total
spin in the process is conserved. Therefore, for a quark with spin parallel to the
total spin of the nucleon $q\uparrow$, we consider the following detailed processes:
\\1. gluon emission: $\gamma^{*}q\uparrow\rightarrow
q\uparrow g$,
\\2. gluon absorption: $q\uparrow g \rightarrow
q\uparrow\gamma^{*}$,
\\3. pair production: $\gamma^{*}g\rightarrow
q\uparrow q\downarrow$,
\\4. pair annihilation: $q\uparrow q\downarrow
\rightarrow \gamma^{*} g$,
\\while for $q\downarrow$, they are:
\\1.gluon emission: $\gamma^{*}q\downarrow\rightarrow q\downarrow g$,
\\2. gluon absorption: $q\downarrow g \rightarrow q\downarrow\gamma^{*}$,
\\3. pair production: $\gamma^{*}g\rightarrow q\uparrow
q\downarrow$,
\\4. pair annihilation: $q\uparrow q\downarrow \rightarrow \gamma^{*}g$.

Thus the cross sections of the above mentioned processes should be added into the
parton model cross section. So the modified quark distribution functions could be
written as
\begin{eqnarray}
q\uparrow(x, Q^{2})=q\uparrow(x) +\frac{\alpha_{s}}{2\pi}\log
\frac{Q^{2}}{\mu^{2}}{\int}_{x}^{1}\frac{dy}{y}q\uparrow(y)P_{q\uparrow
q\uparrow}(\frac{x}{y}) \nonumber\\ \nonumber
 +\frac{\alpha_{s}}{2\pi}\log
 \frac{Q^{2}}{\mu^{2}}{\int}_{x}^{1}\frac{dy}{y}g(y)P_{q\uparrow g}(\frac{x}{y}) \\  \nonumber
 -\frac{\alpha_{s}}{2\pi}\log \frac{Q^{2}}{\mu^{2}}{\int}_{0}^{x}\frac{dy}{x}q\uparrow(x)P_{q\uparrow
q\uparrow} (\frac{y}{x})\\
 -\frac{\alpha_{s}}{2\pi}\log \frac{Q^{2}}{\mu^{2}}{\int}_{0}^{x}\frac{dy}{x}g(x)P_{q\uparrow
g} (\frac{y}{x})
\end{eqnarray}
for $q\uparrow$, where $\mu$ is a cutoff value of the transverse momentum, and
\begin{eqnarray}
q\downarrow(x, Q^{2})=q\downarrow(x)+\frac{\alpha_{s}}{2\pi}\log
\frac{Q^{2}}{\mu^{2}}{\int}_{x}^{1}\frac{dy}{y}q\downarrow(y)P_{q\downarrow
q\downarrow} (\frac{x}{y}) \nonumber \\ \nonumber
 +\frac{\alpha_{s}}{2\pi}\log \frac{Q^{2}}{\mu^{2}}{\int}_{x}^{1}\frac{dy}{y}g(y)P_{q\downarrow
g} (\frac{x}{y})\\ \nonumber
 -\frac{\alpha_{s}}{2\pi}\log \frac{Q^{2}}{\mu^{2}}{\int}_{0}^{x}\frac{dy}{x}q\downarrow(x)P_{q\downarrow
q\downarrow} (\frac{y}{x}) \\
 -\frac{\alpha_{s}}{2\pi}\log \frac{Q^{2}}{\mu^{2}}{\int}_{0}^{x}\frac{dy}{x}g(x)P_{q\downarrow
g} (\frac{y}{x})
\end{eqnarray}
 for $q\downarrow$, where $g(x)$ is the gluon
distribution function [2] (here we neglect the gluon evolution for approximation),
\begin{equation}
g(x)=\frac{16M^{2}Rx}{2\pi}{\int}_{\frac{Mx}{2}}^{\frac{M}{2}} \frac{1}
{\exp(\frac{E}{T})-1} \frac{dE}{E^{2}}.
\end{equation}

$P_{q\uparrow q\uparrow}, P_{q\downarrow q\downarrow}$ are the splitting functions
denoting the probability of emission of a gluon with a fraction of the momentum of
the parent quark which emits it, and $P_{q\uparrow g}, P_{q\downarrow g}$ are the
splitting functions denoting the probability of emission of a quark with a fraction
of the momentum of the parent gluon. Thus the second to fifth items in eq (2), (3)
denote the contribution of the gluon emission, pair production, gluon absorption and
pair annihilation respectively. From the calculating of the cross section of the each
process [18], we have
\begin{equation}
P_{q\uparrow q\uparrow}(z)=P_{q\downarrow q\downarrow}(z)=
\frac{4}{3}\frac{1+z^{2}}{(1-z)}_{+}+ 2 \delta(1-z)
\end{equation}
and
\begin{equation}
P_{q\uparrow g}(z)=P_{q\downarrow g}(z)=\frac{1}{2}(z^{2}+(1-z)^{2}).
\end{equation}
\\
Substitute eq (1), (4), (5), (6) into eq (2), (3) we get:
\begin{equation}
\frac{2\pi}{\alpha_{s}}\frac{d q\uparrow(x, Q^{2})}{d \log Q^{2}}=
2q\uparrow(x)-\frac{1}{3}g(x)+ \frac{3M^{2}Rx}{2\pi}{\int}_{x}^{1} f\uparrow(z)dz,
\end{equation}
where \[
 f\uparrow(z)= \frac{4}{3}\frac{1+z^2}{(1-z)z^2}
{\int}_{\frac{Mx}{2z}}^{\frac{M}{2}}\frac{1} {\exp(\frac{E-{\mu}_{q\uparrow}}{T})+1}
\frac{dE}{E^2} -\frac{8}{3(1-z)}{\int}_{\frac{Mx}{2}}^{\frac{M}{2}} \frac{1}
{\exp(\frac{E-{\mu}_{q\uparrow}}{T})+1} \frac{dE}{E^2}
\] \[
+\frac{8}{3}\frac{z^2+(1-z)^2}{z^2}{\int}_{\frac{Mx}{2z}}^{\frac{M}{2}} \frac{1}
{\exp(\frac{E}{T})-1} \frac{dE}{E^2},
\]
and
\begin{equation}
\frac{2\pi}{\alpha_{s}}\frac{d q\downarrow(x, Q^{2})}{d \log Q^{2}}=
2q\downarrow(x)-\frac{1}{3}g(x)+ \frac{3M^{2}Rx}{2\pi}{\int}_{x}^{1} f\downarrow(z)dz,
\end{equation}
where \[
 f\downarrow(z)= \frac{4}{3}\frac{1+z^2}{(1-z)z^2}
{\int}_{\frac{Mx}{2z}}^{\frac{M}{2}}\frac{1}
{\exp(\frac{E-{\mu}_{q\downarrow}}{T})+1} \frac{dE}{E^2}
-\frac{8}{3(1-z)}{\int}_{\frac{Mx}{2}}^{\frac{M}{2}} \frac{1}
{\exp(\frac{E-{\mu}_{q\downarrow}}{T})+1} \frac{dE}{E^2}
\] \[
+\frac{8}{3}\frac{z^2+(1-z)^2}{z^2}{\int}_{\frac{Mx}{2z}}^{\frac{M}{2}} \frac{1}
{\exp(\frac{E}{T})-1} \frac{dE}{E^2},
\]
in which, the quark's potential could be written as [2]
\begin{equation}
\mu_{q\uparrow\downarrow}=\frac{\pi}{6R}(v_{q}\pm \Delta_{q}),
\end{equation}
where $v_{q}$ is the valence quark number of flavor $q$ and $\Delta_{q}$ is the
number difference between the quark with spin parallel and anti-parallel to the total
spin of flavor $q$.

Eq (7) and (8) cannot be integrated analytically except in the region $x\rightarrow
1$ or $x\rightarrow 0$. At x=0.03, 0.06, 0.1, 0.2, 0.3 ,0.4 ,0.5 ,0.6, 0.7, We have
calculated $q(x)$ and $\frac{2\pi}{\alpha_{s}}\frac{d q(x, Q^{2})}{d \log Q^{2}}$
numerically for each quark
($u\uparrow,u\downarrow,\bar{u}\uparrow,\bar{u}\downarrow,d\uparrow,d\downarrow,
\bar{d}\uparrow,\bar{d}\downarrow,s\uparrow,s\downarrow,\bar{s}\uparrow,\bar{s}\downarrow$)
inside a proton. We use the parameters that $T=63 MeV$, $R=(180MeV)^{-1}$, $M=938
MeV$, $\Delta_{u}=0.83$, $\Delta_{d}=-0.43$, $\Delta_{s}=-0.10$ [2, 19, 20, 21] for
the proton. The results are shown in Table 1 and Table 2.

From Table 1 and 2, we can get some conclusions:
\\1. $\frac{2\pi}{\alpha_{s}}\frac{d q(x, Q^{2})}{d \log Q^{2}}$ are
quite the same at very small $x$ for all quarks and at very large $x$ for the quarks
with small potentials.
\\2. $\frac{2\pi}{\alpha_{s}}\frac{d q(x, Q^{2})}{d \log Q^{2}}$
change from positive to negative at about $x=0.3\sim 0.4$ for dominant quarks (the
quarks with big potentials) and at $x$ around 0.5 for the quarks with small
potentials.

The structure function could be calculated by [18], [22]
\begin{equation}
F_{2}(x, Q^{2})=x {\sum_{q \bar{q} \uparrow\downarrow}} e_{q}^{2}\left[
q(x)+\frac{\alpha_{s}}{2\pi}\log \frac{Q^2}{{\mu}^2}\left(2q(x)-\frac{1}{3}g(x)+
\frac{3M^{2}Rx}{2\pi}{\int}_{x}^{1}f(z)dz \right)\right],
\end{equation}
where $f(z)$ is defined by eq (7), (8). The running coupling constant $\alpha_{s}$
could be written as [23, 24, 25]
\begin{equation}
\alpha_{s}=\frac{4\pi}{9\log \frac{Q^2}{\Lambda^2}},
\end{equation}
where $\Lambda^2$ is a renormalization constant. Substitute eq (11) into eq (10), we
get the $\alpha_{s}$ dependent structure function:
\begin{equation}
F_{2}(x, Q^{2})=x {\sum_{q \bar{q} \uparrow\downarrow}} e_{q}^{2}\left[
q(x)+\frac{2}{9}\left( 1-\frac{\alpha_{s}}{\alpha_{s}({\mu}^2)} \right)
\left(2q(x)-\frac{1}{3}g(x)+ \frac{3M^{2}Rx}{2\pi}{\int}_{x}^{1}f(z)dz \right)\right],
\end{equation}
where $\alpha_{s}({\mu}^2)$ is the coupling constance at $Q^{2}={\mu}^2$.

We have plotted $F_{2}(x, Q^{2})\sim x $ at $\alpha_{s}=\alpha_{s}({\mu}^2)$,
$\alpha_{s}=0.99\alpha_{s}({\mu}^2) $, $\alpha_{s}=0.9 \alpha_{s}({\mu}^2)$,
$\alpha_{s}=0.5 \alpha_{s}({\mu}^2)$ (see Figure 1, 2, 3, 4) and $F_{2}(x, Q^{2})\sim
\alpha_{s}$ at $x=0.03, x=0.1, x=0.2, x=0.3, x=0.4, x=0.5, x=0.6, x=0.7$ (see Figure
5) for the proton, which shows an agreement with the experiments [26].

\begin{table} [1]
\caption{$q(x)\sim x$ of each quark inside the proton ($T=63 MeV$, $R=(180MeV)^{-1}$,
$M=938 MeV$). }
\begin{center}
\begin{tabular}{|c|c|c|c|c|c|c|c|c|c|}\hline
 \hline   &$u\uparrow$ &$u\downarrow$ &$d\uparrow$ &$d\downarrow$ &$s\uparrow$
 &$s\downarrow$\\
\hline $\mu$(MeV)  &267  &110&54&135&-9.4&9.4\\
\hline x=0.03 & 2.85 &2.15&1.60&2.34&0.937&1.13\\
\hline x=0.06  &2.68 &1.81&1.27&2.03&0.686&0.842\\
\hline x=0.1 &2.44&1.43 &0.925&1.65&0.462&0.579\\
\hline x=0.2  &1.84&0.756 &0.414&0.939&0.179&0.233\\
\hline x=0.3  &1.31 &0.369&0.179&0.490&0.0714&0.0945\\
\hline x=0.4  &0.864 &0.169&0.0750&0.235&0.0289&0.0387\\
\hline x=0.5  &0.516&0.0741&0.0318&0.106&0.0119&0.0159\\
\hline x=0.6  &0.276&0.0316&0.0133&0.0462&0.00490&0.00660\\
\hline x=0.7  &0.133&0.0131&0.00546&0.0194&0.00200&0.00270\\
\hline   &$\bar{u}\downarrow$ &$\bar{u}\uparrow$
&$\bar{d}\downarrow$&$\bar{d}\uparrow$  &$\bar{s}\downarrow$&$\bar{s}\uparrow$   \\
\hline $\mu$(MeV)  & -267 &-110&-54&-135&9.4&-9.4\\
\hline x=0.03 &0.0241&0.264&0.562&0.184&1.13&0.937\\
\hline x=0.06 &0.0160&0.180&0.395&0.124&0.842&0.686\\
\hline x=0.1  &0.00988&0.113&0.256&0.0776&0.579&0.462\\
\hline x=0.2  &0.00337&0.0398&0.0935&0.0270&0.233&0.179\\
\hline x=0.3  &0.00126&0.0151&0.0362&0.0102&0.0945&0.0714 \\
\hline x=0.4  &0.000497&0.00598&0.0144&0.00403&0.0387&0.0289\\
\hline x=0.5  &0.000201&0.00243&0.00589&0.00164&0.0159&0.0119\\
\hline x=0.6  &0.000083&0.000998&0.00243&0.000671&0.00660&0.00490\\
\hline x=0.7  &0.000034&0.000407&0.000989&0.000274&0.00270&0.00200\\
\hline
\end{tabular}
\end{center}
\end{table}

\begin{table} [2]
\caption{$\frac{2\pi}{\alpha_{s}}\frac{d q(x, Q^{2})}{d \log Q^{2}}\sim x$ of each
quark inside the proton ($T=63 MeV$, $R=(180MeV)^{-1}$, $M=938 MeV$). }
\begin{center}
\begin{tabular}{|c|c|c|c|c|c|c|c|c|c|}\hline
 \hline   &$u\uparrow$ &$u\downarrow$ &$d\uparrow$ &$d\downarrow$ &$s\uparrow$
 &$s\downarrow$\\
\hline  $\mu$(MeV)  &267  &110&54&135&-9.4&9.4\\
\hline x=0.03 &14812 &14813&14813&14813&14814&14814\\
\hline x=0.06  &1408 &1409&1409&1409&1410&1410\\
\hline x=0.1 &205&206 &206&206&207&207\\
\hline x=0.2  &8.26&8.98 &9.21&8.86&9.36&9.33\\
\hline x=0.3  &0.0908 &0.717&0.843&0.637&0.914&0.899\\
\hline x=0.4  &-0.451 &0.00955&0.0712&-0.0343&0.102&0.0958\\
\hline x=0.5  &-0.329&-0.0365&-0.00854&-0.0578&0.00466&0.00197\\
\hline x=0.6  &-0.184&-0.0222&-0.0102&-0.0319&-0.00462&-0.00574\\
\hline x=0.7  &-0.0893&-0.0106&-0.00554&-0.0147&-0.00327&-0.00373\\
\hline   &$\bar{u}\downarrow$ &$\bar{u}\uparrow$
&$\bar{d}\downarrow$&$\bar{d}\uparrow$  &$\bar{s}\downarrow$&$\bar{s}\uparrow$   \\
\hline $\mu$(MeV)  & -267 &-110&-54&-135&9.4&-9.4\\
\hline x=0.03 &14815&14814&14814&14815&14814&14814\\
\hline x=0.06 &1410&1410&1410&1410&1410&1410\\
\hline x=0.1  &207&207&207&207&207&207\\
\hline x=0.2  &9.48&9.46&9.42&9.47&9.33&9.36\\
\hline x=0.3  &0.961&0.951&0.937&0.955&0.899&0.914 \\
\hline x=0.4  &0.121&0.117&0.112&0.119&0.0958&0.102\\
\hline x=0.5  &0.0124&0.0109&0.00862&0.0114&0.00197&0.00466\\
\hline x=0.6  &-0.00144&-0.00205&-0.00299&-0.00183&-0.00574&-0.00462\\
\hline x=0.7  &-0.00197&-0.00222&-0.00260&-0.00213&-0.00373&-0.00327\\
\hline
\end{tabular}
\end{center}
\end{table}

\begin{figure}[1]
\begin{center}
\leavevmode \epsfxsize=15cm\epsfbox{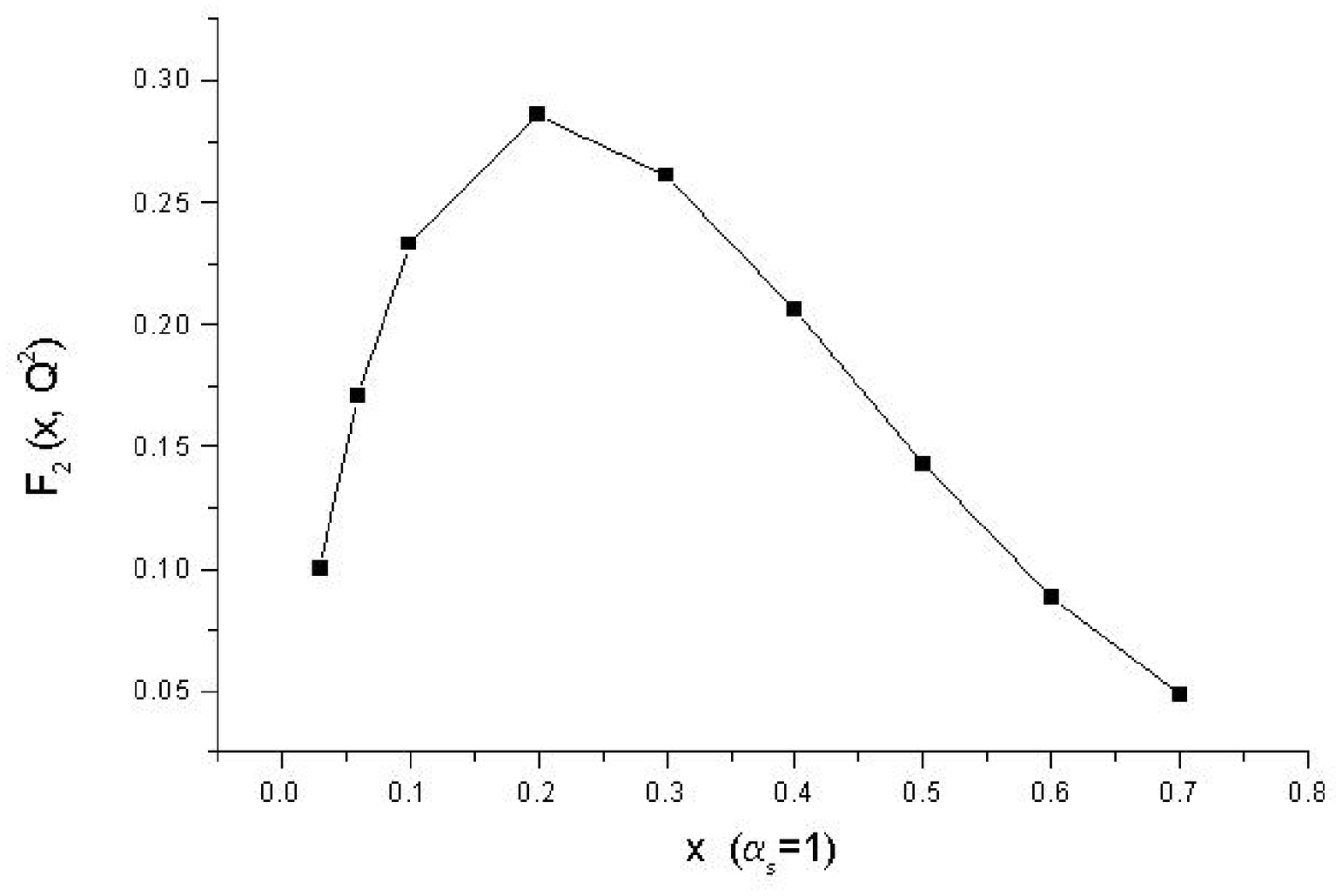}
\end{center}
\caption{ $F_{2}(x, Q^{2})\sim x $ of the proton at $\alpha_{s}=\alpha_{s}({\mu}^2)$.
The unit of $\alpha_{s}$ is $\alpha_{s}({\mu}^2)$.}
\end{figure}

\begin{figure}[2]
\begin{center}
\leavevmode \epsfxsize=15cm\epsfbox{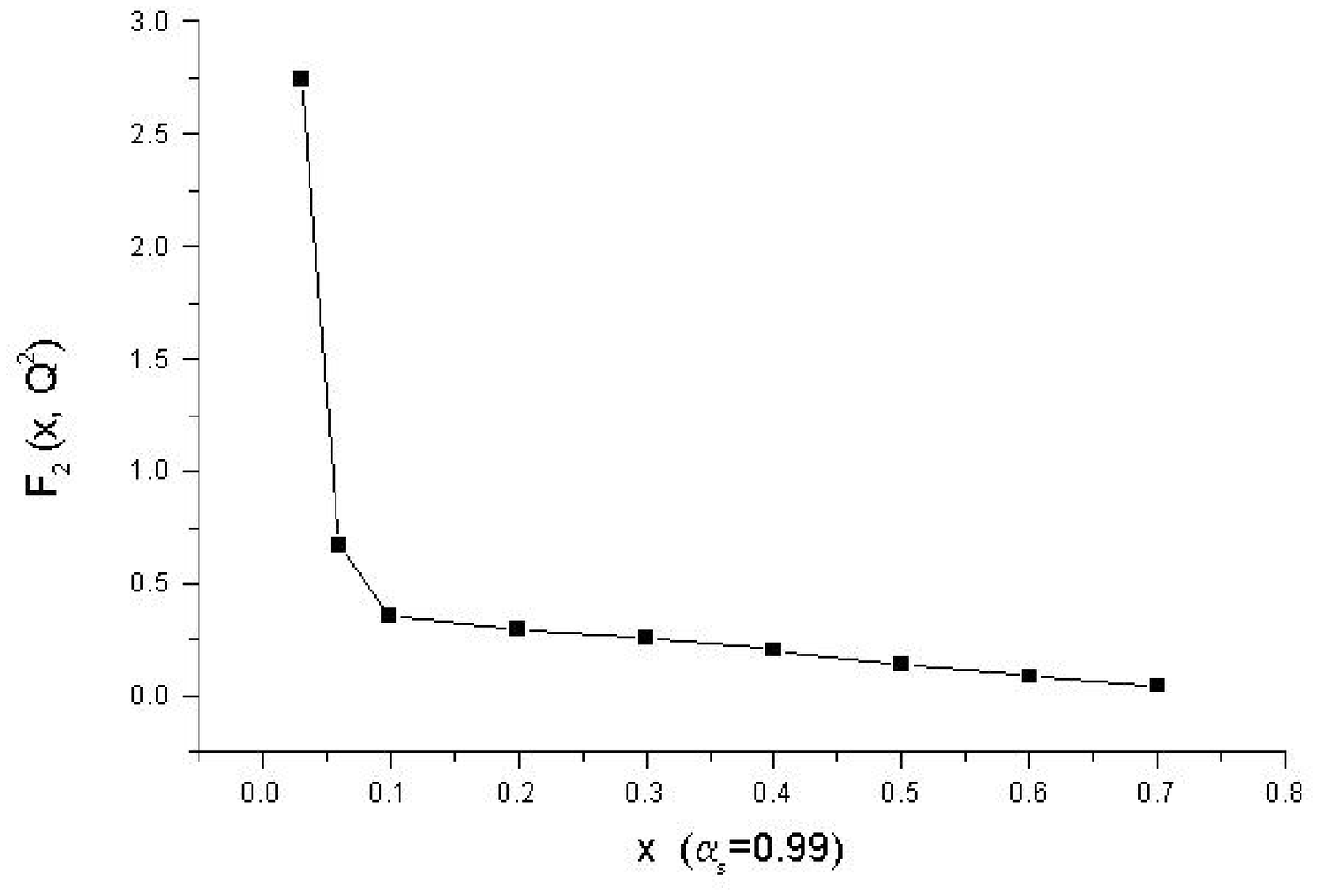}
\end{center}
\caption{ $F_{2}(x, Q^{2})\sim x $ of the proton at $\alpha_{s}=0.99
\alpha_{s}({\mu}^2)$. The unit of $\alpha_{s}$ is $\alpha_{s}({\mu}^2)$.}
\end{figure}

\begin{figure}[3]
\begin{center}
\leavevmode \epsfxsize=15cm\epsfbox{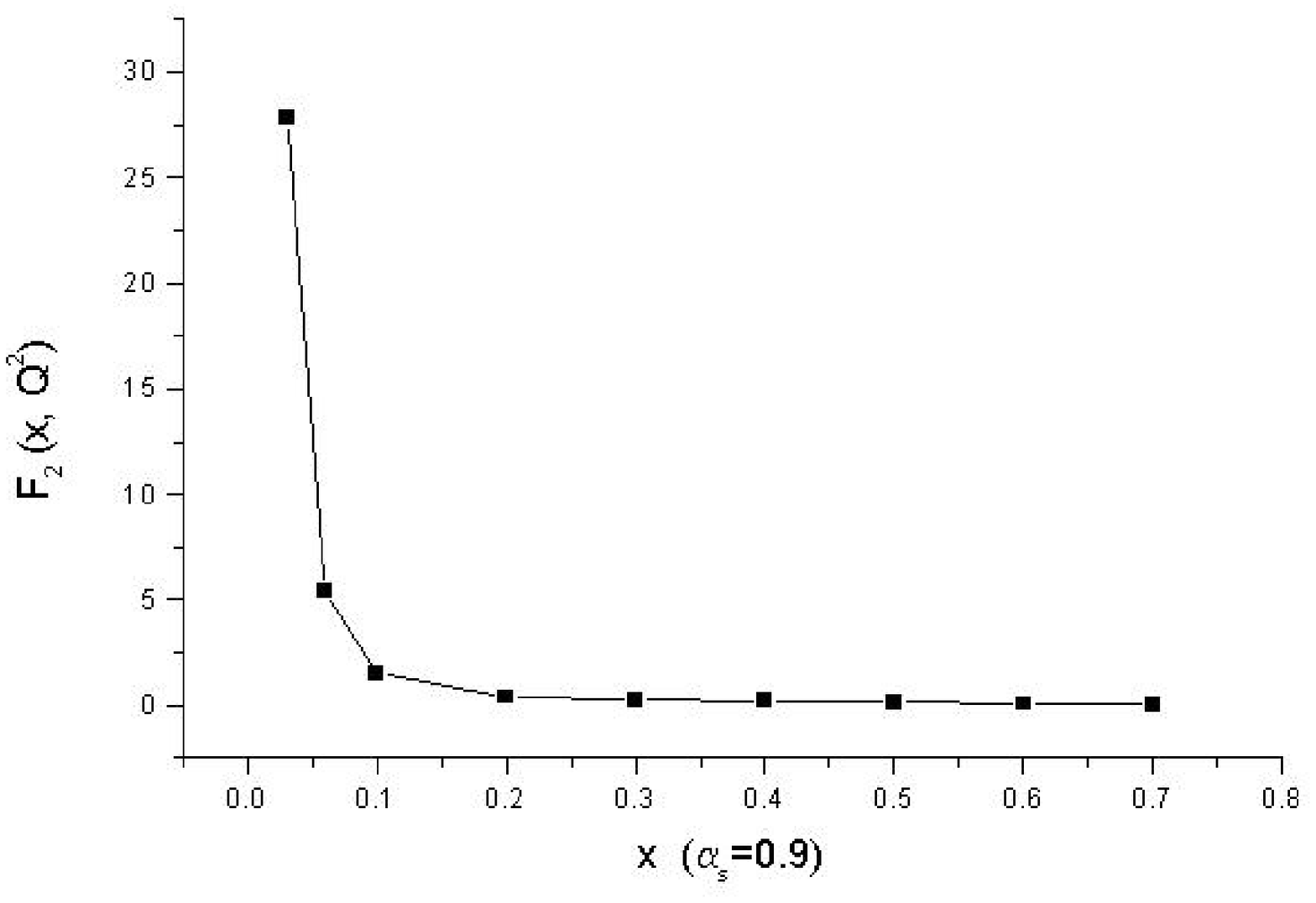}
\end{center}
\caption{ $F_{2}(x, Q^{2})\sim x $ of the proton at $\alpha_{s}=0.9
\alpha_{s}({\mu}^2)$. The unit of $\alpha_{s}$ is $\alpha_{s}({\mu}^2)$.}
\end{figure}

\begin{figure}[4]
\begin{center}
\leavevmode \epsfxsize=15cm\epsfbox{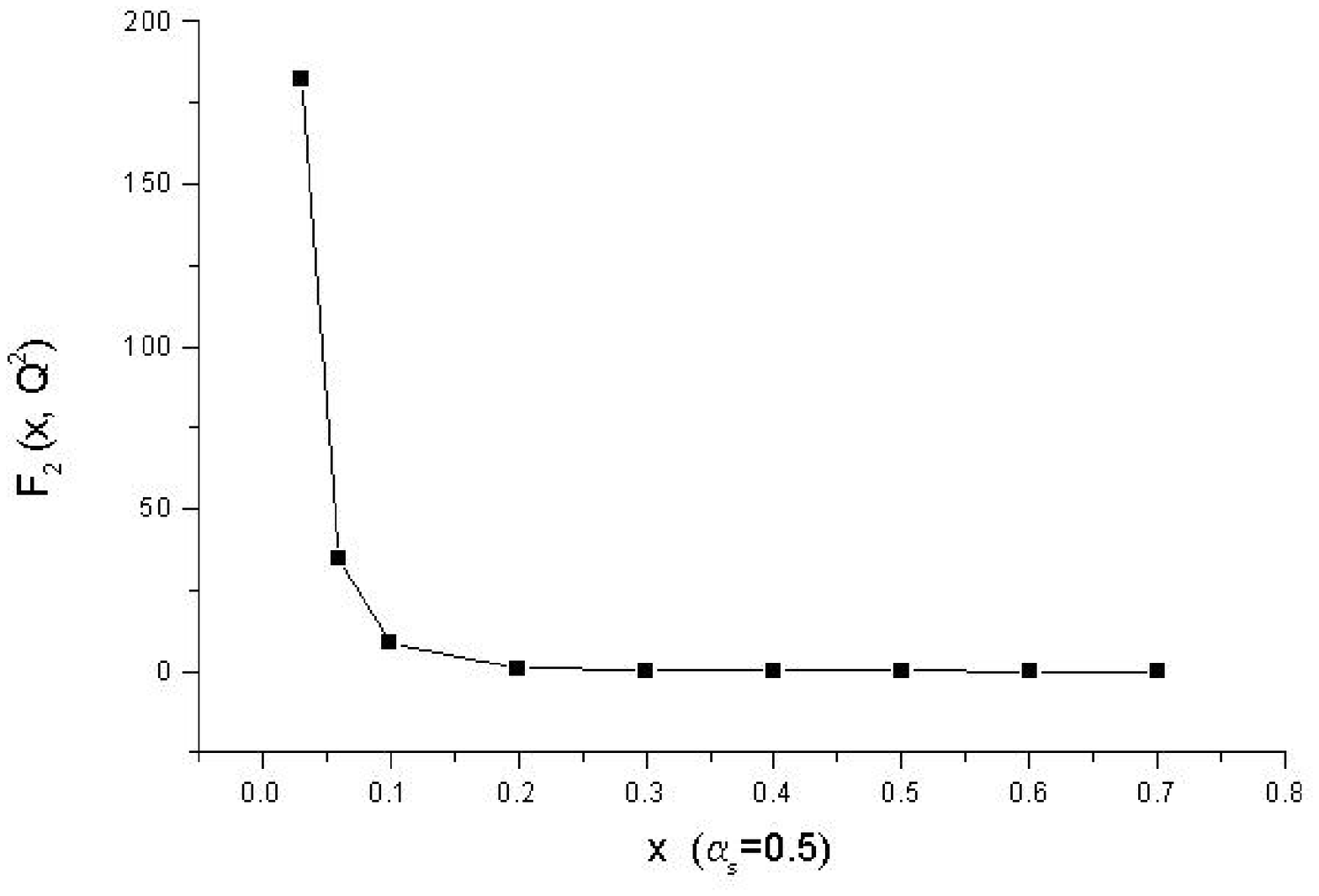}
\end{center}
\caption{ $F_{2}(x, Q^{2})\sim x $ of the proton at $\alpha_{s}=0.5
\alpha_{s}({\mu}^2)$. The unit of $\alpha_{s}$ is $\alpha_{s}({\mu}^2)$.}
\end{figure}

\begin{figure}[5]
\begin{center}
\leavevmode \epsfxsize=15cm\epsfbox{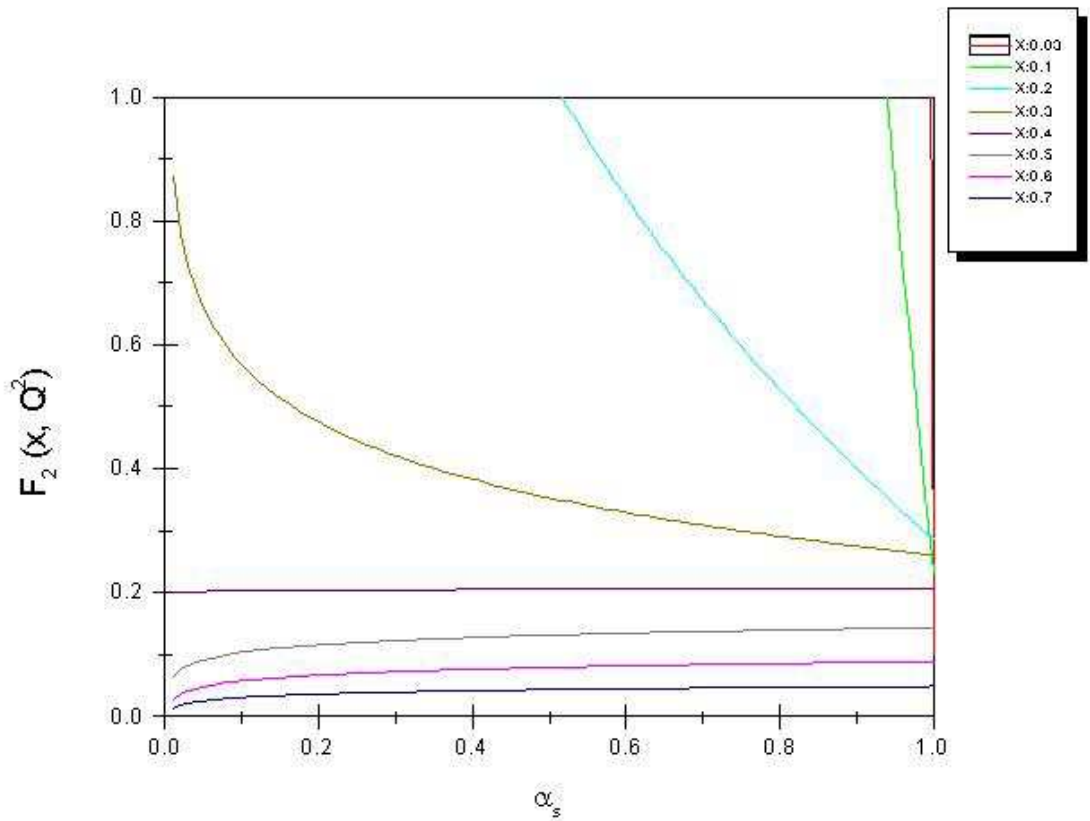}
\end{center}
\caption{ $F_{2}(x, Q^{2})\sim \alpha_{s}$ of the proton at $x=0.03, x=0.1, x=0.2,
x=0.3, x=0.4, x=0.5, x=0.6, x=0.7$. The unit of $\alpha_{s}$ is
$\alpha_{s}({\mu}^2)$.}
\end{figure}

\end{document}